\documentclass[]{PoS}
\usepackage{amsmath} 

\title{On the reduction of Feynman integrals to master integrals}
\author{\speaker{A.V.~Smirnov}$^a$ and V.A.~Smirnov$^b$\\
\llap{$^a$} Scientific Research Computing Center of Moscow State University, Moscow, Russia\\
\llap{$^b$} Nuclear Physics Institute of Moscow State University, Moscow, Russia\\
E-mail: \email{asmirnov@rdm.ru},
\email{smirnov@theory.sinp.msu.ru}}

\abstract{The reduction of Feynman integrals to master integrals
is an algebraic problem that requires algorithmic approaches at
the modern level of calculations. Straightforward applications of
the classical Buchberger algorithm to construct Gr\"obner bases
seem to be inefficient. An essential modification
designed especially for this problem has been worked out. It has
been already applied in two- and three-loop calculations.
We are also suggesting to combine our method with the Laporta's algorithm.}

\FullConference{XI International Workshop on Advanced Computing and Analysis Techniques in Physics Research
\\ Amsterdam, Holland\\ April, 23-27 2007}
\ShortTitle{Reduction of Feynman integrals}

\newcommand{\bea}{\begin{eqnarray}}
\newcommand{\eea}{\end{eqnarray}}
\newcommand{\dd}{\mbox{d}}

\begin{document}

\section{Introduction. Reduction problem for Feynman integrals}

After a tensor reduction is performed
a given Feynman graph $\Gamma$
results into various scalar Feynman
integrals that have the same structure of the integrand with
various distributions of powers of propagators
\bea
  F(a_1,\ldots,a_n) =
  \int \cdots \int \frac{\dd^d k_1\ldots \dd^d k_h}
  {E_1^{a_1}\ldots E_n^{a_n}}\,.
  \label{eqbn}
\eea
$d=4-2\epsilon$; the denominators $E_r$ are either quadratic or linear with respect
to the loop momenta $p_i=k_i, \; i=1,\ldots,h$ and
to the independent external momenta $p_{h+1}=q_1,\ldots,p_{h+N}=q_N$ of the graph.

There are different methods of evaluating those integrals. An old analytical
strategy is to evaluate, by some methods, every scalar Feynman integral
generated by the given graph.
And what has already become a traditional strategy is to derive, without
calculation, and then apply integration by parts (IBP) relations \cite{IBP}
between the given family of Feynman integrals as recurrence relations.

After such a reduction a general integral from the family (\ref{eqbn}) is expressed as a linear
combination of some basic ({\em master}) integrals.
Thus, the whole problem of evaluation falls apart into two parts:
constructing a reduction procedure
and evaluating master integrals.
This paper is devoted to the first part of the problem.

Let us introduce some notation.
Let ${\mathcal F}=F(a_1,\ldots,a_n)$ be functions of integer variables
$a_1,\ldots,a_n$; it is an infinitely dimensional linear
space. The simplest basis consists of elements
\bea
H_{a_1,\ldots,a_n}(a'_1,\ldots,a'_n)=\delta_{a_1,a'_1}\ldots\delta_{a_n,a'_n}.
\eea

From this point of view, Feynman integrals form a point in ${\mathcal F}$;
all relations we use turn into linear functionals on this vector space.

There are different types of relations. The most commonly used among them are
IBP relations \cite{IBP}
\bea
\int\ldots\int \dd^d k_1 \dd^d k_2\ldots
\frac{\partial}{\partial k_i}\left( p_j
\frac{1}{E_1^{a_1}\ldots E_n^{a_n}}
\right)   =0  \;,
\eea

After differentiation we obtain some relations of the following
form:
\bea
\sum \alpha_i F(a_1+b_{i,1},\ldots,a_n+b_{i,n}) =0\,.
\label{IBP}
\eea

There are more relations one can consider:
Lorentz-invariance (LI) identities \cite{GR1},
symmetry relations, e.g.,
\bea
F(a_1,\ldots,a_n)=(-1)^{d_1 a_1+\ldots d_n
a_n}F(a_{\sigma(1)},\ldots,a_{\sigma(n)}),
\eea
boundary conditions: i.e. the conditions of the following form:
\bea
F(a_1,a_2,\ldots,a_n)=0\mbox{ when }a_{i_1}\leq 0,\ldots a_{i_k}\leq 0
\label{boundary}
\eea
for some subsets of indices $i_j$;
parity conditions and others.
All those relations lead to a possibility to express given Feynman
integrals in terms of other Feynman integrals.

Therefore one has to name certain integrals irreducible ({\em master}) and aim to
reduce any other integral to those.
An attempt to formalize the definition of master integrals was
made in \cite{ourwork2}.
To define master integrals one needs to introduce a priority
between the points $(a_1,\ldots,a_n)$, hence {\em an ordering}.

There are different ways to choose an ordering. To start with,
Feynman integrals are simpler, from the analytic point of view, if they have more
non-positive indices. Moreover, solving IBP relations by hand
usually consists of reducing indices to zero.

Naturally, we come to the notion of {\em sectors}.
Those are $2^n$ regions labeled by subsets
$\nu \subseteq \{1,\ldots,n\}$, where
$\sigma_{\nu}$ = $\{ (a_1,\ldots,a_n): a_i>0\;\;
\mbox{if}\;\; i\in \nu\,,\;\; a_i\leq 0\;\;
\mbox{if}\;\; i \not\in \nu\}$.
A sector $\sigma_{\nu}$ is said to be {\em lower} than a sector
$\sigma_{\mu}$ if $\nu\subset\mu$.
Furthermore, $F(a_1,\ldots,a_n)\succ F(a'_1,\ldots,a'_n)$
if the sector of $(a'_1,\ldots,a'_n)$ is lower than the sector of
$(a_1,\ldots,a_n)$.
To define an ordering completely
one has to introduce it in some way inside the sectors (this will be discussed
below).

Initially they used to solve relations by hand, but with the
growth of the complexity of the problems it turned to be
impractical. The first algorithmic approach to solving IBP relations
was the so-called Laporta's algorithm \cite{Lap1,Lap2,Lap3,Lap4} which is now
very well known and actively used in practice.
It is based on the fact that  the total number
of IBP  equations written for concrete indices grows faster than the number of
independent Feynman integrals, hence this system of equations sooner or later
becomes overdetermined, and one obtains a possibility to perform
a reduction to master integrals. At the moment, there is one public implementation
of this algorithm called {\tt AIR} \cite{AnLa} and a lot of private
versions.\footnote{by T.~Gehrmann and E.~Remiddi, M.~Czakon,
P.~Marquard and D.~Seidel, Y.~Schr\"oder, C.~Sturm, A.~Onishchenko, O.~Veretin, \ldots}

There are other algorithmic approaches to solving the relations
such as the Baikov's method
\cite{Bai1,Bai2,Bai3,Bai4} and the use of Gr\"obner bases \cite{Buch}.
The reduction using Gr\"obner bases has been suggested by Tarasov \cite{Tar1,Tar2}
who reduced the reduction problem to differential equations by
introducing a mass for every line,
$a_i {\bf i}^+\to \frac{\partial}{\partial m^2_i}$.
A direct application of Gr\"obner bases (without the use of differential equations)
was initially suggested by Gerdt \cite{Gerdt}.

One more approach initially based on Gr\"obner bases,
called the $s$-bases, has been developed in \cite{ourwork} (see also \cite{ourwork22}
for a brief review). It was applied in practice for
a reduction of a family of three-loop Feynman integrals necessary for the
analysis of decoupling of $c$-quark loops in $b$-quark HQET \cite{G2S}.
In the next section we outline basic ideas of this
approach in an updated form. Then, in Section~3, we formulate how it can be
combined with other strategies, in particular with the Laporta's algorithm.

\section{The $s$-bases approach}

Suppose first that we are interested in expressing any
integral in the positive sector
$\sigma_{\{1,\ldots,n\}}$ as a linear combination of a finite number
of integrals in it.
The left-hand sides of IBP relations (\ref{IBP}) can be expressed in terms of
operators of multiplication $A_i$ and shift
operators $Y_i={\bf i}^+,Y^-_i={\bf i}^-$, where
\bea
(A\cdot F)(a_1,\ldots,a_n)&=&a_i F(a_1,\ldots,a_n)\;\mbox{ and}
\\
(Y^{\pm}_i\cdot F)(a_1,\ldots,a_n)&=&F(a_1,\ldots, a_i\pm 1,\ldots,a_n)\;.
\eea

Let ${\cal A}_{1,\ldots,n}$ be the algebra generated by shift operators $Y^+_i$ and
multiplication operators $A_i$. It acts on the field of functions
$\mathcal F$ of $n$ integer variables.
One can choose certain elements $f_i$ corresponding to IBP
relations and write
\bea
(f_i \cdot F)(a_1,\ldots,a_n) =0\;.
\label{ideal}
\eea

The choice of elements $f_i$ is not unique,
we will choose them so that they do not depend on $Y^-_i$.
Consider the left ideal
${\cal I}\subset {\cal A}_{1,\ldots,n}$, generated by the elements $f_i$.
This ideal is named as the ideal of IBP relations.
For any element $X\in\cal I$ we have
\bea
(X F)(1,1,\ldots,1) = 0 \;.
\eea

Also we have
\bea
F(a_1,a_2,\ldots,a_n)=({Y_1^{a_1-1}\ldots Y_n^{a_n-1}} F)(1,1,\ldots,1).
\label{initial}
\eea

The idea of the algorithm is to reduce the operator on the
right-hand side of (\ref{initial}) using the elements of the ideal
$\cal I$. Suppose we are interested in $F(a_1,a_2,\ldots,a_n)$.
The reduction problem becomes equivalent to
reducing the monomial $Y_1^{a_1-1}\ldots Y_n^{a_n-1}$
modulo the ideal of the IBP relations. After obtaining an
expression like
\bea
Y_1^{a_1-1}\ldots Y_n^{a_n-1}= \sum r_i f_i
+ \sum c_{i_1,\ldots, i_n} Y_1^{i_1-1}\ldots Y_n^{i_n-1}
\label{reduction}
\eea
it is left to apply (\ref{reduction}) to $F$ at $a_1=1,\ldots,a_n=1$
and obtain the following expression:
\bea
F(a_1,a_2,\ldots,a_n)=\sum c_{i_1,\ldots, i_n}
F(i_1,i_2,\ldots,i_n)\,,
\eea
where integrals on the right-hand side are ``master integrals''
(the formulas (\ref{ideal}) and (\ref{initial}) are used).

To do the reduction we need an ordering of monomials of operators
$Y_i$ or, similarly, an ordering of points $(a_1,\ldots,a_n)$ in
the sector: (for two monomials $M_1=Y_1^{i_1-1}\ldots
Y_n^{i_n-1}$ and $M_2=Y_1^{j_1-1}\ldots Y_n^{j_n-1}$
we have
$(M_1 \cdot F)(1,\ldots,1) \succ (M_2 \cdot F)(1,\ldots,1)$
if and only if $M_1 \succ M_2$).
Then the reduction procedure becomes similar to the division of
polynomials. But one needs to introduce a proper ordering.

A monomial is defined by its degree, i.e. a set of $n$ non-negative integers ($\mathbb N^n$).
Thus defining an ordering on monomials is equivalent to defining
an ordering on $\mathbb N^n$. We require the following properties:

i) for any $a\in \mathbb N^n$ not equal to $(0,\ldots 0)$ one has $(0,\ldots 0)\prec a$;
\\
\indent ii) for any $a,b,c\in \mathbb N^n$ one has $a\prec b$
if and only if $a+c\prec b+c$.

For example, a lexicographical ordering is defined the following way:
a set $(i_1,\ldots,i_n)$ is said to be higher than a set
$(j_1,\ldots,j_n)$ (denoted by $(i_1,\ldots,i_n)\succ
(j_1,\ldots,j_n)$) if there is $l\leq n$ such
that $i_1=j_1$, $i_2=j_2$, \ldots, $i_{l-1}=j_{l-1}$ and
$i_l>j_l$.

Another example is a degree-lexicographical ordering:
$(i_1,\ldots,i_n)\succ (j_1,\ldots,j_n)$
if $\sum i_k >   \sum j_k$, or $\sum i_k =   \sum j_k$
and $(i_1,\ldots,i_n)\succ (j_1,\ldots,j_n)$ in the sense of
the lexicographical ordering.

An ordering can be defined by a non-degenerate $n\times n$ matrix ($a_{k,l}$):
for two sets of numbers $(i_1,\ldots,i_n)$ and $(j_1,\ldots,j_n)$
one first compares $\sum_l i_l a_{1,l}$ and
$\sum_l j_l a_{1,l}$. If the first number is greater, then the
first degree is greater; if the first number is smaller, then the
first degree is smaller; and if those
numbers are equal we compare $\sum_l i_l a_{2,l}$ and
$\sum_l j_l a_{2,l}$ and so on.

The following matrices correspond to a lexicographical, a degree-lexicographical and a reverse
degree-lexicographical ordering for $n=5$:

\vspace{2mm}

$\begin{pmatrix}
  1 & 0 & 0 & 0 & 0 \\
 0 & 1 & 0 & 0 & 0 \\
 0 & 0 & 1 & 0 & 0 \\
  0 &0 & 0 & 1 & 0 \\
 0 & 0 & 0 & 0 & 1
\end{pmatrix},\;\;\;\;
\begin{pmatrix}
  1 & 1 & 1 & 1 & 1 \\
 1 & 0 & 0 & 0 & 0 \\
 0 & 1 & 0 & 0 & 0 \\
  0 &0 & 1 & 0 & 0 \\
 0 & 0 & 0 & 1 & 0
\end{pmatrix},\;\;\;\;
\begin{pmatrix}
  1 & 1 & 1 & 1 & 1 \\
 1 & 1 & 1 & 1 & 0 \\
 1 & 1 & 1 & 0 & 0 \\
  1 &1 & 0 & 0 & 0 \\
 1 & 0 & 0 & 0 & 0
\end{pmatrix}$

\vspace{2mm}

Such an approach encounters the following problem:
the reduction does not always lead to a reasonable number of irreducible
integrals, so one has to build a special basis of the ideal first.
Obviously having elements with lowest possible degrees
is equivalent to obtaining master integrals with minimal possible
degrees. Therefore one needs to build special bases.
This can be done by an algorithm based
on the Buchberger algorithm with the use of $S$-polynomials and
reductions \cite{Buch}.

Moreover, one must keep in mind that we are interested in integrals not only in the
positive sector.
Our algorithm \cite{ourwork} aims to build a set of special bases
of the ideal of IBP relations ($s$-bases).
The basic idea is to consider consider the algebra
${\cal A}_\nu$ generated
by operators $A_i$ and operators $Y_i^+$ for $i\in\nu$ and
$Y_i^-$ for $i\not\in\nu$.
Then for $\sigma_{\nu}$ we again consider
the ideal of IBP relations in ${\cal A}_\nu$.
Now one has to construct \textit{sector bases} ($s$-bases), rather than
Gr\"obner bases for all sectors, where an $s$-basis for a sector $\sigma_{\nu}$
is a set of elements of $\cal I$ which provides the possibility
of a reduction to master integrals in $\sigma_{\nu}$ and integrals whose
indices lie in lower sectors, i.e. $\sigma_{\nu'}$ for $\nu'\subset\nu$.

This leads to considering many sectors --- seemingly the problem becomes harder. But the
important simplification is that one is not trying to solve the
reduction problem in each sector separately but allows to reduce
the integrals in a given sector to lower sectors --- similarly to
the ``by hand'' solutions. It is also worth noting that
it is most complicated
to construct $s$-bases for minimal sectors.

The construction of $s$-bases is close to the Buchberger algorithm but it can be terminated
when the ``current'' basis already provides us the needed reduction.
The basic operations are the same,
i.e. calculating $S$-polynomials and
reducing them modulo current basis, with a chosen ordering.

After constructing $s$-bases for all non-trivial sectors one
obtains a recursive (with respect to the sectors)
procedure to evaluate $F(a_1,\ldots,a_n)$ at any point and
thereby reduce a given integral to master integrals.
A description of the old version of the algorithm (implemented in {\tt Mathematica}) can be found in \cite{mypaper}.

\section{Combining $s$-bases with other strategies}

Constructing a basis requires a certain skill to find a proper
ordering. At a high level (starting from 10 to 12 indices) it turns to be a
certain problem. In some sectors the ordering can be found
automatically, in some sectors it is easily done manually,
but in some sectors the $s$-bases can be hardly constructed.
There are several ideas that might allow the algorithm to
construct the bases better and we hope that they will be
implemented in the future.
Currently this is not a problem
of cpu time or memory, but an algorithmical one: with a fixed
ordering the algorithm either produces a basis quickly, or does
not do it at all.


Recently we have worked out one more way to improve the algorithm.
Experience shows that  $s$-bases are constructed easily if the number of
non-positive indices in a given sector is small.
And what can help is the fact that if the number of non-positive indices
is large, there is usually a possibility to perform
integration over some loop momentum
explicitly in terms of gamma functions for general $d$.
However,
to derive the corresponding explicit formulae, with multiple
finite summations, for all necessary cases turns out to be
impractical.

In this situation, there is another alternative. Let us
distinguish the propagators (and irreducible numerators) involved
in such an explicit integration formula. They are associated with a
subgraph of the given graph. Let us solve IBP relations for the
corresponding subintegral in order to express any such subintegral
in term of some master integrals.

Then, after using this reduction
procedure, it will be sufficient to use explicit integration
formulae only for some boundary values of the indices. This
replacement is very simple, without multiple summations.
Coefficients in these reduction formulae depend not only on $d$
and given external parameters but also on the propagators of the
given graph, which are external for the subgraph.

It turned out possible to implement the solution of the
recursive problem for the subgraph in terms of the Feynman
integrals for the whole graph. In this reduction, pure powers
of the parameters which are external for the subgraph
transform naturally into the corresponding shift operators and their
inverse.
Integrals which are obtained from initial integrals by an explicit
integration over a loop momentum in terms of gamma function
usually involve a propagator with an analytic regularization by an amount
proportional to $\epsilon$. Our code is applicable also in such
situations.

Even if we have not constructed the $s$-bases for all sectors we
can still run the reduction procedure. However, in some sectors
there will be an infinite number of irreducible integrals left.

One of the ideas is to use the Laporta's algorithm in those sectors.
Still it leads to several problems in linking two algorithms
together. One of the disadvantages is that the reduction works
much slower if the number of irreducible integrals goes high.

Another idea is to combine two
algorithms inside one code. The $s$-bases might give the
reductions in all sectors, where a basis
can be constructed. And for the
remaining sectors the systems of initial IBP's
are used like in the Laporta's algorithm.

Let us summarize the ideas of this paper. The $s$-bases approach seems to be a good addition to the
methods of automatic integral reduction. The bases construction part of the algorithm should still be
improved, but even now it is capable of constructing bases for
quite complicated cases. A constructed basis in a sector means that in this sector one
needs no more real solving of equations --- the system of linear
equations becomes upper-triangle.

Moreover, it looks like the future is in combining the $s$-bases and the
Laporta's algorithm. An algorithm named {\tt FIRE} (Feynman Integral
REduction) combining all mentioned ideas is currently being
developed. We plan on performing much more complicated three-loop
calculations with the use of {\tt FIRE} (together with M.~Steinhauser; the project is in progress).
Afterwards we intend to make the algorithm public.

{\em Acknowledgements.}
The work was supported by the Russian Foundation for Basic Research through grant
05-02-17645.

\end{document}